\documentstyle[preprint,aps]{revtex}

\begin{document}

\draft

\title{
Semileptonic decays, magnetic moments and spin distributions of spin-1/2 baryons
with sea contribution
}

\author{
V.~Gupta$^a$,\ 
R.~Huerta$^a$,\ 
and
G.~S\'anchez-Col\'on$^{a,b}$
}
\address{
$^a$Departamento de F\'{\i}sica Aplicada. \\
Centro de Investigaci\'on y de Estudios Avanzados del IPN.
Unidad M\'erida. \\ 
A.P. 73, C.P. 97310. Cordemex, M\'erida, Yucat\'an, MEXICO. \\
}
\address{
$^b$Physics Department, University of California Riverside. \\ 
Riverside, CA 92521-0413, U.S.A. \\
}

\date{December 17, 1997}

\maketitle

\begin{abstract}
Spin-1/2 baryons are considered as a composite system made out of a \lq\lq core"
of three quarks surrounded by a \lq\lq sea" (of gluons and $q\bar{q}$-pairs)
which is specified by its total quantum numbers.
Specifically, we assume this sea to be a flavor octet with spin-0 or 1 but no
color.
We show our model can provide very goods fits to magnetic moments and
semileptonic decay data {\em using experimental errors}.
The predictions for spin distributions are in reasonable agreement with
experiment.
\end{abstract}


\section{Introduction}
\label{i}

The expectation of the standard quark model (SQM) that the valence quarks give
dominant contribution to the low energy properties of the spin 1/2 baryons has
had limited quantitative succes.
Recent experiments show that the valence quarks cannot even account for the
proton spin\cite{1}.
One must go beyond SQM.

In reality quarks interact and one expects the physical hadrons to consist of
valence quarks surrounded by a \lq\lq sea" which in general contains gluons and
virtual quark-antiquark ($q\bar{q}$) pairs.
Different treatments of the sea can be found in the
literature\cite{2,3,4,5,6,7}.
We model the general sea by its total quantum numbers (flavor, spin, and color)
which are such that the sea wavefunction when combined with the valence quark
wavefunction gives the desired quantum numbers for the physical hadron.

In particular, the spin 1/2 baryons are pictured as a composite system made out
of a baryon \lq\lq core" of the three valence quarks (as in SQM) and a flavor
octet sea with spin 0 and 1 but no color.
The physical baryon wavefunction incorporating such a \lq\lq sea" was used by
us to calculate the baryon magnetic moments\cite{8}.
Very good fits to the magnetic moment data using {\em experimental errors } were
obtained.
The purpose of this paper is to apply this wavefunction to other low energy
properties of the spin 1/2 baryons $(p,n,\Lambda,\dots)$ like semileptonic
decays (SLD) and baryon spin distributions.

In Sec.~\ref{ii} we give the wavefunction for the physical baryons constructed
from the valence quarks and our model for the sea.
Sec.~\ref{iii} summarizes the results for magnetic moments obtained
earlier\cite{8}.
Sec.~\ref{iv} discusses the application to $G_A$ and $G_V$ for SLD.
Sec.~\ref{v} gives the results for combined fits to 8 magnetic moment data and
the 4 $G_A/G_V$ data.
Sec.~\ref{vi} discusses implications of these fits for the nucleon spin
distribution data.
Sec.~\ref{vii} gives a brief summary.

\section{Spin 1/2 baryon wavefunction with sea}
\label{ii}

The general physical baryon given below in Eq.~(\ref{dos}) was given earlier in
Ref.~\cite{8}.
We give its basic details here not only to establish our notation but also to
make this paper self-contained.

We assume the core baryon wavefunction to be given by the SQM.
For the $SU(3)$ flavor octet spin 1/2 baryons we denote the SQM or $q^3$
wavefunction by $\tilde{B}({\bf 8},1/2)$.
These octet states are denoted by $\tilde{p}$, $\tilde{\Sigma}^{+}$,
$\tilde{\Lambda}$, etc.
The sea is assumed to be a color singlet but with spin and flavor.
Its $SU(3)$ flavor singlet and octet wavefunctions are denoted by $S(\bf 1)$
and $S(\bf 8)$.
These can carry spin 0 (wavefunction $H_0$) or spin 1 (wavefunction $H_1$).
In our model the general sea is described effectively by the four wavefunctions 
$S({\bf 1})H_{0}$, $S({\bf 8})H_{0}$, $S({\bf 1})H_{1}$, and $S({\bf 8})H_{1}$.
We refer to the even parity spin 0 (spin 1) sea as scalar (vector) sea.
The $SU(3)$ symmetric and spinless sea component implicit in SQM is described by
$S({\bf 1})H_{0}$.

The total flavor-spin wavefunction of a spin up ($\uparrow$) physical baryon,
which consists of 3 valence quarks and a sea component as above, can be written
schematically as

\begin{eqnarray}
B(1/2\uparrow)
&=&
\tilde{B}({\bf 8},1/2\uparrow)H_{0}S({\bf 1})+
b_{{\bf 0}}\left[\tilde{B}({\bf 8},1/2)
\otimes H_{1}\right]^{\uparrow}S({\bf 1})
\nonumber\\
&&
+\sum_{N}a(N)\left[\tilde{B}({\bf 8},1/2\uparrow)H_{0}
\otimes S({\bf 8})\right]_{N}
\label{dos}\\
&&
+\sum_{N}b(N)\left\{[\tilde{B}({\bf 8},1/2)\otimes H_{1}]^{\uparrow}
\otimes S({\bf 8})\right\}_{N}.
\nonumber
\end{eqnarray}

\noindent
The normalization not indicated here is discussed later.
The first term is the usual $q^{3}$-wavefunction of the SQM and the second term
(coefficient $b_0$) comes from a flavored singlet vector (spin-1) sea which
combines with the spin 1/2 core baryon $\tilde{B}$ to form a spin 1/2$\uparrow$
state.
So that,

\begin{equation}
\left[\tilde{B}({\bf 8},1/2)\otimes H_{1}\right]^{\uparrow} = 
\sqrt{\frac{2}{3}}\tilde{B}({\bf 8},1/2\downarrow)H_{1,1} -
\sqrt{\frac{1}{3}}\tilde{B}({\bf 8},1/2\uparrow)H_{1,0}.
\label{tres}
\end{equation}

\noindent
The third (fourth) term in Eq.~(\ref{dos}) contains a scalar (vector) sea
which transforms as a flavor octet.
The $SU(3)$ flavor representations obtained from
$\tilde{B}({\bf 8})\otimes S({\bf 8})$ are labelled by
$N={\bf 1,8_{F},8_{D},10,\bar{10},27}$.
The color wavefunctions have not been indicated since the three valence quarks
in the core $\tilde{B}$ and the sea (by assumption) are in a color singlet
state.

As it stands, Eq.~(\ref{dos}) represents a spin 1/2$\uparrow$ baryon which is
not {\em a pure flavor octet} but has an admixture of other $SU(3)$
representations determined by the parameters $a(N)$ and $b(N)$ for
$N={\bf 1,10,\bar{10},27}$.
However, our wavefunction respects isospin ($I$) and hypercharge ($Y$), so that
the physical baryon $B$ ($p$, $n$, $\Lambda$, etc.) have the usual $I$ and $Y$
properties.

The sea isospin multiplets contained in the octet $S({\bf 8})$ are denoted as

\begin{equation}
(S_{\pi^+},S_{\pi^0},S_{\pi^-}),\qquad
(S_{K^+},S_{K^0}),\qquad
(S_{\bar{K}^0},S_{K^-}),\quad
\hbox{and}\quad S_{\eta}.
\label{cuatro}
\end{equation}     
 
\noindent
The suffix on the components label the isospin and hypercharge quantum numbers.
Note, the familiar pseudoscalar mesons are used here as subscripts only to
label the flavor quantum numbers of the sea states.
All the components of $S({\bf 8})$ have $J^P=0^+$ or $1^+$ as mentioned earlier.
For example, $S_{\pi^+}$ has $I=1,I_3=1$ and $Y=0$; $S_{K^-}$ has
$I=1/2,I_3=-1/2$, and $Y=-1$; etc.
These flavor quantum numbers when combined with those of the three valence
quarks states $\tilde{B}$ will give the observed $I$, $I_3$, and $Y$ for the
physical states $B$.
The flavor combinations in the third and fourth terms in Eq.~(\ref{dos}) imply
that the
physical states $B(Y,I,I_3)$ are expressed as a sum of products of
$\tilde{B}(Y,I,I_3)$ and the sea components $S(Y,I,I_3)$, weighted by some
coefficients $\alpha_i$ which are linear combinations of the coefficients
$a(N)$ and $b(N)$.
Schematically, the flavor content of the third and fourth terms in
Eq.~(\ref{dos}) is of the form (suppressing $I_3$)

\begin{equation}
B(Y,I)=
\sum_i \alpha_i(Y_1,Y_2,I_1,I_2)\left[\tilde{B}(Y_1,I_1)S(Y_2,I_2)\right]_i
\label{cinco}
\end{equation}

\noindent
where the sum is over all $Y_i$, $I_i$, ($i=1,2$); such that: $Y=Y_1+Y_2$ and 
$\bf I=I_1+I_2$. 
The flavor content of $B(Y,I,I_3)$ in terms of $\tilde{B}(Y,I,I_3)$ and sea
components are given in Table~\ref{tabla1}.
The corresponding coefficients $\bar{\beta}_i$, $\beta_i$, etc. expressed in
terms of the coefficients $a(N)$ (for the scalar sea) are recorded in
Table~\ref{tabla2}.
In Table~\ref{tabla1} we have denoted $\tilde{B}(Y,I,I_3)$ and $S(Y,I,I_3)$ by
appropiate symbols, e.g., $\tilde{B}(1,1,1/2)$ by $\tilde{p}$, $S(0,1,1)$ by
$S_{\pi^+}$, etc.
Since the flavor content of the fourth term with vector sea is the same as for
the scalar sea, the contribution of the fourth term in Eq.~(\ref{dos}) to the
physical baryon state can be obtained by using Eq.~(\ref{tres}) and
Tables~\ref{tabla1} and \ref{tabla2} with the replacement
$a(N)\rightarrow b(N)$ for $N={\bf 1,8_F,8_D,10,\bar{10},27}$.
For later use, the coefficients obtained by changing $a(N)\rightarrow b(N)$ in
$\bar{\beta}_i$, $\beta_i$, $\gamma_i$, and $\delta_i$ will be denoted by
$\bar{\beta}'_i$, $\beta'_i$, $\gamma'_i$, and $\delta'_i$.
In Tables~\ref{tabla1} and \ref{tabla2} for the reduction of
$\tilde{B}({\bf 8})\otimes S({\bf 8})$ into various $SU(3)$ representations we
have followed the convention used by Carruthers~\cite{carruthers}.

The normalization of the physical baryons wavefunction in Eq.~(\ref{dos}) can
be obtained by using
$\langle H_i | H_j\rangle  = \delta_{ij}$,
$\langle \tilde{B}(Y,I,I_3)|\tilde{B}(Y',I',I'_3)\rangle =
\langle S(Y,I,I_3)|S(Y',I',I'_3)\rangle =
\delta_{YY'}\delta_{II'}\delta_{I_3I'_3}$.
However, it should be noted that the normalization are different, in general,
for each $B(Y,I)$ state.
This is because not all $a(N)$ and $b(N)$ contribute to a given
$(Y,I)$-multiplet as is clear from Tables~\ref{tabla1} and \ref{tabla2}.
For example, $a({\bf 1})$ and $b({\bf 1})$ contribute only to $\Lambda$ while
$a({\bf 10})$ and $b({\bf 10})$ do not contribute to the nucleon states.
Denoting by $N_1$, $N_2$, $N_3$, and $N_4$, the normalization constants for
the $(p,n)$, $(\Xi^0,\Xi^-)$, $(\Sigma^{\pm},\Sigma^0)$, and $\Lambda$
isospin multiplets, one has

\begin{mathletters}
\label{seis}
\begin{equation}
N^2_1 = N^2_0 + a^2({\bf\bar{10}}) + b^2({\bf\bar{10}}),
\end{equation}
\begin{equation}
N^2_2 = N^2_0 + a^2({\bf 10}) + b^2({\bf10}),
\end{equation}
\begin{equation}
N^2_3 = N^2_0 + \sum_{N={\bf 10,\bar{10}}}[a^2(N) + b^2(N)],
\end{equation}
\begin{equation}
N^2_4 = N^2_0 + a^2({\bf 1}) + b^2({\bf 1}),
\end{equation}

\noindent
where,

\begin{equation}
N^2_0 = 1 + b^2_0 + \sum_{N={\bf 8_D,8_F,27}}[a^2(N) + b^2(N)].
\end{equation}
\end{mathletters}

\noindent
For example, using Tables~\ref{tabla1} and \ref{tabla2}, and Eqs.~(\ref{seis}),
the physical spin-up proton state as given by Eq.~(\ref{dos}) is

\begin{eqnarray}
N_1 |p\uparrow\rangle 
&=&
|\tilde{p}\uparrow\rangle H_0S({\bf 1})+b_0|
(\tilde{p}\otimes H_1)^{\uparrow}\rangle S({\bf 1}) 
\nonumber\\
&& 
+ \bar{\beta}_1 |\tilde{p}\uparrow\rangle  S_{\eta} 
+ \bar{\beta}_2 |\tilde{\Lambda}\uparrow\rangle  S_{K^+}
+ \bar{\beta}_3 |(\tilde{N}\uparrow S_{\pi})_{1/2,1/2}\rangle 
+ \bar{\beta}_4 |(\tilde{\Sigma}\uparrow S_K)_{1/2,1/2}\rangle 
\label{siete}\\
&&
+ \bar{\beta}'_1 |(\tilde{p}\otimes H_1)^{\uparrow}\rangle  S_{\eta}
+ \bar{\beta}'_2 |(\tilde{\Lambda}\otimes H_1)^{\uparrow}\rangle  S_{K^+}
\nonumber\\
&&
+ \bar{\beta}'_3 |((\tilde{N}\otimes H_1)^{\uparrow} S_{\pi})_{1/2,1/2}\rangle 
+ \bar{\beta}'_4 |((\tilde{\Sigma}\otimes H_1)^{\uparrow} S_K)_{1/2,1/2}\rangle 
\nonumber,
\end{eqnarray}

\noindent
where $(\tilde{B}\otimes H_1)^{\uparrow}$ are given in Eq.~(\ref{tres}) and
$\bar{\beta}'_1=
[3b({\bf 27})-b({\bf 8_D})+(b({\bf 8_F})+b({\bf \bar{10}}))/2]/\sqrt{20}$,
and so on.
Other baryon wavefunctions will have a similar structure.
Also, $(\tilde{N}\uparrow S_{\pi})_{1/2,1/2}$
($(\tilde{\Sigma}\uparrow S_{K})_{1/2,1/2}$) stand for the $I=I_3=1/2$
combination of the $I=1/2$ $\tilde{N}$ ($S_K$) and $I=1$ $S_{\pi}$
($\tilde{\Sigma}$) multiplets.

For any operator $\hat{O}$ which depends only on quarks, the matrix elements
are easily obtained using the ortogonality of the sea components.
Clearly $\langle p\uparrow|\hat{O}|p\uparrow\rangle $ will be a linear
combination of the matrix elements
$\langle \tilde{B}\uparrow|\hat{O}|\tilde{B'}\uparrow\rangle$ (known from
SQM) with coefficients which depend on the coefficients in the wavefunction.

For applications, we need the quantities $(\Delta q)^B$, $q=u,d,s$; for each
spin-up baryon $B$.
These are defined as 

\begin{equation}
(\Delta q)^B =
n^B(q\uparrow)-n^B(q\downarrow)+n^B(\bar{q}\uparrow)-n^B(\bar{q}\downarrow),
\label{ocho}
\end{equation}

\noindent
where $n^B(q\uparrow)$ ($n^B(q\downarrow)$) are the number of spin-up
(spin-down) quarks of flavor $q$ in the spin-up baryon $B$.
Also, $n^B(\bar{q}\uparrow)$ and $n^B(\bar{q}\downarrow)$ have a similar
meaning for antiquarks.
However, these are zero as there are no explicit antiquarks in the
wavefunctions given by Eq.~(\ref{dos}).
The expressions for $(\Delta q)^B$ are given in Table~\ref{tabla3} in terms of
the coefficients $b_0$, $\bar{\beta}_i$, $\bar{\beta}'_i$, etc.
Note that the terms involving $b_0$, $\bar{\beta}'_i$, $\beta'_i$,
$\gamma'_i$, and $\delta'_i$ are multiplied by the factor $-1/3$ which comes
from Eq.~(\ref{tres}) on taking the matrix element of the operator
$\hat{\Delta q}$.
The expressions for $(\Delta q)^B$ reduce to the SQM values if there is no sea
contribution, that is, $b_0=0$, $a(N)=b(N)=0$,
$N={\bf 1,8_F,8_D,10,\bar{10},27}$.
Moreover, the total spin $S_z$ of a baryon is given by
$S^B_z = (1/2) \sum_q (\Delta q)^B + (\Delta(\hbox{sea}))^B$,
where the second term represents the spin carried by the sea and depends
solely on $b_0$ and $b(N)$'s, the coefficients determining the vector sea.
For $S^B_z=1/2$, we expect $\sum_q (\Delta q)^B = 1$ for a purely scalar sea,
i.e., when $b_0$ and all $b(N)$'s are zero.
This is indeed true for each baryon as can be seen from Table~\ref{tabla3}.
There are three $(\Delta q)^B$ $(q=u,d,s)$ for each $(Y,I)$-multiplet.
These twelve quantities and $(\Delta q)^{\Sigma^0\Lambda}$ are given in terms
of the thirteen parameters of Eq.~(\ref{dos}) as our spin 1/2 baryons do not
belong to a definite representation of $SU(3)$.
To obtain a flavor octet physical baryon one restricts $N$ to ${\bf 8_F}$ and
${\bf 8_D}$ in Eq.~(\ref{dos}), that is, put $a(N)=b(N)=0$ for
$N={\bf 27,10,\bar{10},1}$, so that the twelve $(\Delta q)^B$ are given in
terms of five parameters $b_0$, $a(N)$, $b(N)$ with $N={\bf 8_F,8_D}$.

In our wavefunction, each physical baryon is represented as a superposition of
different combinations of the core baryons with the appropiate sea states
resulting in a very different quark content for the physical baryon state
$B(Y,I,I_3)$.
For example, the physical proton $p$ will contain terms involving $\tilde{p}$,
$\tilde{\Lambda}$, $\tilde{\Sigma^+}$, etc. plus sea and can have non-zero
strange quark content unlike in SQM.
In our model, the baryon $B$ is part of the time just $\tilde{B}$ with an inert
sea (first term in Eq.~(\ref{dos})) and part of the time $\tilde{B}$ plus a sea
with flavor and spin.

How do the sea wavefunctions with $J^P=0^+$ or $1^+$ and the above flavor
properties arise?
One way a sea with flavor ${\bf 8}$ property can arise is from Goldstone bosons
(usual $J^P=0^-$ pseudoscalar mesons, $\pi^{\pm}$, $K^{\pm}$, etc.). Their
effect on baryon structure has been considered recently\cite{eichten}.
These can combine with $q\bar{q}$-pairs or gluons to give the total quantum
numbers for the sea considered by us.

Our approach can be used to construct wavefunctions for other hadrons
incorporating a sea specified by total quantum numbers.
Also, since we have an explicit wavefunction we can calculate all relevant
physical quantities in terms of the parameters in the wavefunction, namely,
$b_0$, $a(N)$'s, and $b(N)$'s.
Since, there is no \'a priori theoretical knowledge which of these are
important, we determine them by confronting the predictions of our wavefunctions
with experiment.

\section{Application to magnetic moments}
\label{iii}

For this purpose, the baryon magnetic moment operator $\hat{\mu}$ was assumed to
be expressed solely in terms of valence quarks (in $\tilde{B}$)  so that
$\hat{\mu}=\sum_q (e_q/2m_q) \sigma^q_Z$ ($q=u,d,s$).
In principle, we could consider a magnetic moment operator $\hat{\mu}^{(s)}$
for the `sea' and through which the vector sea would contribute to
$\mu_B$.
We do not include such a direct sea contribution as it would
involve unknown parameters like $\mu_{S_{\pi^+}}$, $\mu_{S_{K^-}}$, etc.
Furthermore, the sea is specified by its total quantum numbers.
Since a given set of quantum numbers can be achieved by a multitude of
different configurations of $q\bar{q}$-pairs and gluons, one may assume that
the overall contribution due to $\hat{\mu}^{(s)}$ is negligible.
It is clear from Eq.~(\ref{dos}) that $\mu_B=\langle B|\hat{\mu}|B\rangle$
will be a linear combination of $\mu_{\tilde{B}}$ and
$\mu_{\tilde{\Sigma}^0\tilde{\Lambda}}$ weighted by coefficients which
depend on $b_0$, $a(N)$'s, and $b(N)$'s.
The $\mu_{\tilde{B}}$ and $\mu_{\tilde{\Sigma}^0\tilde{\Lambda}}$ (for the core
baryons) are given in terms of the quark magnetic moments $\mu_q$ as per SQM.
For example,
$\mu_{\tilde{p}}=(4\mu_u-\mu_d)/3$,
$\mu_{\tilde{\Lambda}}=\mu_s$,
$\mu_{\tilde{\Sigma}^0\tilde{\Lambda}}=(\mu_u-\mu_d)/\sqrt{3}$,
etc.
Consequently, one obtains (for $B=p,n,\Lambda,\ldots$)

\begin{mathletters}
\label{nueve}
\begin{eqnarray}
\mu_B &=& \sum_q (\Delta q)^B \mu_q,\ \ \ \ (q=u,d,s);
\\
\mu_{\Sigma^0\Lambda} &=&
\sum_q (\Delta q)^{\Sigma^0\Lambda} \mu_q,\ \ \ \ (q=u,d);
\end{eqnarray}
\end{mathletters}

\noindent
where the $(\Delta q)^B$ and $(\Delta q)^{\Sigma^0\Lambda}$ are given in
Table~\ref{tabla3}.

For the fits $\mu_q$ (or equivalently $m_q$) were also treated as parameters.
The details of the fits and discussion can be found in Ref.~\cite{8}.
There, {\em using experimental errors }we had obtained two excellent six
parameters fits to the eight magnetic moment data.
The details of these fits (called Case~1 and 2 in Ref.~\cite{8}) are given here
for sake of completeness:

\paragraph*{Case 1.}

The values
$\mu_u=2.5007\mu_N$,
$\mu_d=-1.3058\mu_N$,
$\mu_s=-0.8233\mu_N$,
$a({\bf 8_F})=-0.1536$,
$a({\bf 10})=0.5065$,
and
$b({\bf 8_F})=0.5272$
of the six parameters gave a
$\chi^2/{\rm DOF}=1.95/2$.

\paragraph*{Case 2.}

The values
$\mu_u=2.4748\mu_N$,
$\mu_d=-1.3010\mu_N$,
$\mu_s=-0.8243\mu_N$,
$a({\bf 8_F})=-0.1466$,
$a({\bf 10})=0.4941$,
and
$b_0=0.4779$
of the six parameters gave
$\chi^2/{\rm DOF}=2.09/2$.

A comparison of the two fits shows:
(a) The values of the quark masses satisfy $m_u\approx m_d\approx 0.6 m_s$ in
accord with quark model expectations for both fits.
Also, the masses in the two cases are practically the same.
(b) In either case, the scalar sea is described by the same two parameters
$a({\bf 8_F})$, and $a({\bf 10})$ which have nearly the same values.
c) In both the cases, the vector sea is described by only one parameter but its
nature is very different in the two cases.
In Case~1, the vector sea carries flavor (parameter $b({\bf 8_F})$) while in
Case~2 it is flavorless (parameter $b_0$).
(d) The $SU(3)$ breaking effects in the wavefunction are solely due to the
scalar sea parameter $a({\bf 10})$.
(e) Both fits give $(\Delta u)^p\approx 1$, $(\Delta d)^p\approx -0.25$ with a
small value of $(\Delta s)^p$ with different sign in the two cases: for Case~1,
$(\Delta s)^p\approx -0.009$ while for Case~2, $(\Delta s)^p\approx 0.003$.

The 3-parameter function determined by the magnetic moments can be used to
predict other data.
We consider the predictions for the semileptonic decays below.

\section{Semileptonic decays (SLD)}
\label{iv}

For the semileptonic decay $B\to B'+(\mbox{lepton pair})$ we need to calculate
the matrix elements $G_{V,A}(B\to B')=\langle B'|J_{V,A}|B\rangle$, of the
charged changing hadronic vector ($J_V$) and axial-vector ($J_A$) currents
using our wavefunction.
We discuss this below for both $\Delta S=0$ and $\Delta S=1$ decays separately.

\subsection{${\bf \Delta S=0}$ decays}

These decays at a quark level represent a $d\to u$ transition.

\paragraph*{a) $G_{V}(B\to B')$.}
The vector current $J_V(\Delta S=0)=I_+$ the total isospin raising operator for
the physical baryons.
Since our general wavefunction (in Eq.~(\ref{dos})) respects isospin symmetry,
the allowed $G_{V}(B\to B')$ are easily calculated.
For example, since the physical nucleons $(p,n)$ form a $I=1/2$ doublet,
$G_V(n\to p)=\langle p|J_V(\Delta S=0)|n\rangle=1$ and similarly since $\Lambda$
and $\Sigma^{\pm}$ belong to different isospin representations,
$G_V(\Sigma^{\pm}\to \Lambda)=0$.

Morever, since $I_+$ is a generator of the isospin symmetry it can be written
simply as

\begin{mathletters}
\label{nuevep}
\begin{equation}
I_+=\left(\sum_q I^{(q)}_+\right)+I^{(s)}_+
\label{nuevepa}
\end{equation}

\noindent
where \lq\lq$q$" and \lq\lq$s$" refer to \lq\lq quark" and \lq\lq sea" parts.
The current operator $J^{(q)}_V(\Delta S=0)\equiv\sum_qI^{(q)}_+$ acts on the
quarks in the core baryons $\tilde{B}$ and is the isospin raising operator for
the $\tilde{B}$ states so that
$(\sum_q I^{(q)}_+)|\tilde{n}\rangle=|\tilde{p}\rangle$,
$(\sum_q I^{(q)}_+)|\tilde{\Sigma}^0\rangle=\sqrt{2}|\tilde{\Sigma}^+\rangle$,
etc.
Similarly, $J^{(s)}_V(\Delta S=0)\equiv I^{(s)}_+$ acts on sea states in
$S({\bf 8})$ given in Eq.~(\ref{cuatro}), so that
$I^{(s)}_+|S_{\pi^0}\rangle=\sqrt{2}|S_{\pi^+}\rangle$, etc.\footnote{In our
convention, the isospin multiplets in $\tilde{B}({\bf 8})$ are
$(\tilde{p},\tilde{n})$,
$(\tilde{\Sigma}^+,\tilde{\Sigma}^0,\tilde{\Sigma}^-)$,
$(\tilde{\Xi}^0,\tilde{\Xi}^-)$, and $\tilde{\Lambda}$.
The multiplets in $S({\bf 8})$ are given in Eq.~(\ref{cuatro}).
The raising and lowering operators obey the standard (Condon-Shortley) $SU(2)$
phase conventions.}
It is clear that (\ref{nuevepa}) is equivalent to 

\begin{equation}
J_V(\Delta S=0)=J^{(q)}_V(\Delta S=0)+J^{(s)}_V(\Delta S=0).
\label{nuevepb}
\end{equation}
\end{mathletters}

\noindent
Using Eqs.~(\ref{nuevep}) (and the conventions of Ref.~\cite{carruthers}) one
can explicitly verify that for $\Delta S=0$ decays our wavefunctions in
Eq.~(\ref{dos}) gives the $G_V(B\to B')$ listed in column~2 of
Table~\ref{tabla4n}.

\paragraph*{b) $G_{A}(B\to B')$.}

The total axial current $J_A(\Delta S=0)$ can be written in terms of a quark
part $J^{(q)}_A(\Delta S=0)$ and a sea part $J^{(s)}_A(\Delta S=0)$.
In general,

\begin{equation}
J_A(\Delta S=0)=J^{(q)}_A(\Delta S=0)+A_0J^{(s)}_A(\Delta S=0),
\label{diezn}
\end{equation}

\noindent
where the constant $A_0$ specifies the relative strength of the sea axial
current operator $J^{(s)}_A$ relative to the quark axial current operator
$J^{(q)}_A$.
The reason for the presence of $A_0$ is that $J_A(\Delta S=0)$ is not a
generator of a flavor symmetry as it involves the spin operator.
At present, we have no \'a priori knowledge of the magnitude or sign of $A_0$.

To be able to calculate $G_{A}(B\to B')$ one has to specify $J_A(\Delta S=0)$
completely.
In SQM, one has $J^{(q)}_A(\Delta S=0)=\sum_qI^{(q)}_+\sigma^{(q)}_z$ and in
analogy we take $J^{(s)}_A(\Delta S=0)=2I^{(s)}_+S^{(s)}_z$ where $S^{(s)}_z$
is a spin operator which acts on the sea spin wavefunctions.
Clearly, $S^{(s)}_z|H_{1,m}\rangle=m|H_{1,m}\rangle$, $m=\pm1,0$ and
$S^{(s)}_z|H_0\rangle=0$.
As a result, only the vector sea with flavor will contribute to
$\langle B'|J^{(s)}_A(\Delta S=0)|B\rangle$.

Using these operators, the results for the matrix elements
$G^{(q)}_{A}(B\to B')\equiv\langle B'|J^{(q)}_A(\Delta S=0)|B\rangle$
and
$G^{(s)}_{A}(B\to B')\equiv\langle B'|J^{(s)}_A(\Delta S=0)|B\rangle$
are given in columns 3 and 4 of Table~\ref{tabla4n}.
The full $G_{A}(B\to B')=G^{(q)}_{A}(B\to B')+A_0G^{(s)}_{A}(B\to B')$
is to be used to confront data.

It is worth remarking from Table~\ref{tabla4n} that for transitions within an
isomultiplet, the ratio

\begin{equation}
\frac{G^{(q)}_{A}(B\to B')}{G_V(B\to B')}=(\Delta d)^B-(\Delta d)^{B'}.
\label{oncen}
\end{equation}

\noindent
It is not surprising that Eq.~(\ref{oncen}) (a consequence of isospin for a
$d\to u$ transition) holds because our general wavefunction respects isospin.
Another consequence of isospin (noted here for the first time because of the
parameters in the wavefunction!) is the relation of
$G^{(q)}_{A}(\Sigma^\pm\to\Lambda)$ to
$(\Delta d)^{\Sigma^0\Lambda}\ (=-(\Delta u)^{\Sigma^0\Lambda})$
in Table~\ref{tabla4n}.
The relation in Eq.~(\ref{oncen}) states that $G_A/G_V$ for the quark part in
$B\to B'$ is related to the difference of the spin carried by the decaying
quark in $B$ and $B'$.
We will see below that a similar relationship with $\Delta d$ replaced by
$\Delta s$ holds for the $\Delta S=1$ or $s\to u$ transitions in the limit of
$SU(3)$ symmetry (see Eq.~(\ref{trecen}) below).

\subsection{${\bf \Delta S=1}$ decays}

At the quark level these decays are due to a $s\to u$ transition.
For these decays the current operators can be represented in terms of the
lowering (raising) operators
$V_-$ ($V_+$) of the $V$-spin $SU(2)$ sub-group of flavor $SU(3)$.
In our conventions\cite{carruthers}, $(s,u)$ form a doublet with
$V_-|s\rangle=|u\rangle$.
The corresponding $V$-spin multiplets (with standard $SU(2)$ phase conventions)
in $\tilde{B}({\bf 8})$ are
$(\tilde{\Sigma}^-,\tilde{n})$,
$(\tilde{\Xi}^0,\tilde{\Sigma}^+)$,
$(\tilde{\Xi}^-,(\tilde{\Sigma}^0+\sqrt{3}\tilde{\Lambda})/2,\tilde{p})$,
and
$(\sqrt{3}\tilde{\Sigma}^0-\tilde{\Lambda})/2$,
while those in $S({\bf 8})$ are
$(S_{\pi^-},S_{K^0})$,
$(S_{\bar{K}^0},S_{\pi^+})$,
$(S_{K^-},(S_{\pi^0}+\sqrt{3}S_{\eta})/2,S_{K^+})$,
and
$(\sqrt{3}S_{\pi^0}-S_{\eta})/2$.
The $V$-spin multiplets for the physical baryons can be obtained from
$\tilde{B}({\bf 8})$ by changing $\tilde{B}\to B$.

\paragraph*{a) $G_{V}(B\to B')$.}

The vector current operator $J_V(\Delta S=1)=V_-$ the $V$-spin lowering
operator for the physical baryons.
As it is a flavor generator one can write
$J_V(\Delta S=1)=J^{(q)}_V(\Delta S=1)+J^{(s)}_V(\Delta S=1)$
where
$J^{(q)}_V(\Delta S=1)=\sum_qV^{(q)}_-$
and
$J^{(s)}_V(\Delta S=1)=V^{(s)}_-$ are the $V$-spin lowering operators for
$\tilde{B}({\bf 8})$ and $S({\bf 8})$ states.
Using this one obtains
$G_{V}(B\to B')=\langle B'|J_V(\Delta S=1)|B\rangle$ for $\Delta S=1$ decays
listed in Table~\ref{tabla5n}.
Since $J_V$ does not depend on spin but only flavor the scalar and vector sea
parameters contribute in the same way to $G_{V}(B\to B')$.
Note that one calculates $G_{V}(B\to B')$ initially in terms of the
coefficients $\beta_i$, $\beta'_i$, etc. (see Table~\ref{tabla1})
However, on using Table~\ref{tabla2}, these expressions take a simple form in
terms of the coefficients $a(N)$ and $b(N)$ (in Eq.~(\ref{dos})) and
it is these which are listed in Table~\ref{tabla5n}.
Furthermore, the relations
$\sqrt{2}G_V(\Sigma^0\to p)=G_V(\Sigma^-\to n)$
and
$\sqrt{2}G_V(\Xi^-\to \Sigma^0)=G_V(\Xi^0\to \Sigma^+)$
are due to isospin and can be derived directly.
For example,
$\sqrt{2}\langle p|V_-|\Sigma^0\rangle=\langle p|V_-I_+|\Sigma^-\rangle$
and since $[V_-,I_+]=0$ and $\langle p|I_+=\langle n|$ one has the result.
These isospin relations will clearly hold for the corresponding $G_A$'s.
In the $SU(3)$ limit, when only the coefficients $b_0$, $a(N)$, and $b(N)$
$N={\bf 8_{F}}$ and ${\bf 8_{D}}$ are non-zero in Eq.~(\ref{dos}), the physical
baryons form a $SU(3)$ octet.
This means that $V$-spin is a good symmetry.
Indeed in this limit $G_{V}(B\to B')$ (in Table~\ref{tabla5n}) become numbers
(e.g. $G_{V}(\Lambda\to p)=\sqrt{3/2}$, etc.) as they should.

\paragraph*{b) $G_{A}(B\to B')$.}

Corresponding to Eq.~(\ref{diezn}) in the case of $\Delta S=0$ SLD's one
can write 

\begin{equation}
J_A(\Delta S=1)=J^{(q)}_A(\Delta S=1)+A_1J^{(s)}_A(\Delta S=1),
\label{docen}
\end{equation}

\noindent
in terms of a quark part $J^{(q)}_A(\Delta S=1)$ and a sea part
$J^{(s)}_A(\Delta S=1)$.
The constant $A_1$ specifies the relative strength of sea axial current operator
$J^{(s)}_A(\Delta S=1)$ to the quark axial current operator
$J^{(q)}_A(\Delta S=1)$ since $J_A(\Delta S=1)$ is not a generator.
$A_1$ plays the same role for $\Delta S=1$ decays as $A_0$ does for the
$\Delta S=0$ decays.

To calculate $G_{A}(B\to B')=\langle B'|J_A(\Delta S=1)|B\rangle$ one has to
specify $J^{(q)}_A$ and $J^{(s)}_A$ in Eq.~(\ref{docen}).
Since in SQM $J^{(q)}_A(\Delta S=1)=\sum_qV^{(q)}_-\sigma^{(q)}_Z$, so we take 
$J^{(s)}_A(\Delta S=1)=2V^{(s)}_-S^{(s)}_Z$ as was done for the $\Delta S=0$
decays.
Using these operators the results for the matrix elements
$G^{(q)}_{A}(B\to B')\equiv\langle B'|J^{(q)}_A(\Delta S=1)|B\rangle$
and
$G^{(s)}_{A}(B\to B')\equiv\langle B'|J^{(s)}_A(\Delta S=1)|B\rangle$
are given in Table~\ref{tabla6n}.
It is clear that only sea states with flavor and spin will contribute to
$G^{(s)}_{A}(B\to B')$ (as in $\Delta S=0$ case) so that it ultimately depends
only on the coefficients $b(N)$ in Eq.~(\ref{dos}).

It is clear from Eq.~(\ref{docen}), that
$G_{A}(B\to B')\equiv\langle B'|J_A(\Delta S=1)|B\rangle=
G^{(q)}_{A}(B\to B')+A_1G^{(s)}_{A}(B\to B')$
is to be compared with data.

Some consequences of flavor symmetry need to be pointed out for the expressions
in Table~\ref{tabla6n}.
Firstly, as expected, the explicit calculation confirm that the matrix elements
obey the isospin relations $\sqrt{2}G_A(\Sigma^0\to p)=G_A(\Sigma^-\to n)$ and
$\sqrt{2}G_A(\Xi^-\to \Sigma^0)=G_A(\Xi^0\to \Sigma^+)$.
Furthermore, in the limit when the physical baryons form a $SU(3)$ octet (i.e.
$a(N)=b(N)=0$ for $N={\bf 1,10,\bar{10},27}$ in Eq.~(\ref{dos}))
$G^{(q)}_{A}(B\to B')$ for $\Delta S=1$ decays can be expressed in terms of only
$(\Delta s)^B$ given in Table~\ref{tabla3}.
One has

\begin{equation}
\frac{G^{(q)}_{A}(B\to B')}{G_V(B\to B')}=(\Delta s)^B-(\Delta s)^{B'}.
\label{trecen}
\end{equation}

\noindent
So, for $\Delta S=1$ decays $G_A/G_V$ for the quark part in $B\to B'$ is related
to the difference of the spin carried by the decaying quark in $B$ and $B'$.
This result has not been noted earlier.

In summary, the presence of the sea component in the wavefunction modifies the
expression for $G_V$ and $G_A$ for the $\Delta S=0,1$ decays.
Since our wavefunction respects isospin and the $\Delta S=0$ vector current is
$I_+$, the $G_V(\Delta S=0)$ are just numbers as in SQM.
The $G_V(\Delta S=1)$ have simple dependence on the parameters since the
$\Delta S=1$ vector current is a $V$-spin generator.
The $G_A$'s for $\Delta S=0,1$ SLD are more profoundly changed.
The quark part acquires dependence on the wavefunction parameters because of the
sea components.
In addition, there is a possible {\em direct} sea contribution $G^{(s)}_A$ which
arises only from a vector sea with flavor.
This is weighted by the effective parameters $A_0$ and $A_1$ for $\Delta S=0$
and $\Delta S=1$ decays.
To gauge the effect of this direct sea contribution we attempt fits only with
$A_0,A_1=0$ or $\pm 1$ below.

\section{Fits to the combined magnetic moment and SLD data}
\label{v}

The excellent fits for the magnetic moments given in Sec.~\ref{iii} can
straightaway be used to predict the $G_A/G_V$ for the 4 SLD's $n\to p$,
$\Lambda\to p$, $\Sigma^-\to n$, and $\Xi^-\to\Lambda$ for which data are
available.
A contribution from $G^{(s)}_A$ (direct sea contribution) arises only for
Case~1.
The numerical predictions are poor especially for $n\to p$ and $\Sigma^-\to n$.
The minimum $\chi^2$-fits to magnetic moments alone do not give acceptable fits
to the SLD data.
This situation changes profoundly when a combined fit to {\em both} the 8
magnetic moment data and 4 SLD $G_A/G_V$ data is made with 1 or 2 more
parameters to describe the sea, but reducing one of the $\mu_q$'s as a
parameter, for example, by having $m_u=m_d$.

\subsection{Fits with experimental errors for all data}
\label{va}

The six parameter fits mentioned in Sec.~\ref{iii}\cite{8} to magnetic moment
data describe the sea in terms of only 3 parameters.
Using these values of the 3 sea parameters to predict the SLD $G_A/G_V$ gives at
best a $\chi^2_{\rm SLD}\approx 137$ with $A_0=A_1=0$ for Case~1 and a $\chi^2$
of 76 for Case~2.
However, a combined fit to the SLD and magnetic moment data with 5 parameters to
describe the sea and 2 parameters $\mu_u$ and $\mu_s$ (we put $m_u=m_d$) give an
excellent fit (given in Table~\ref{tabla7n}) with $A_0=0$ and $A_1=-1$, which
determine the strength of the direct sea contribution to $G_A(\Delta S=0)$ and
$G_A(\Delta S=1)$ respectively.
Treating $A_0$ and $A_1$ as free parameters does not affect the $\chi^2$ as
their values come to be $A_0=-0.016$ and $A_1=-1.01$.
Thus our combined fit gives $\chi^2_{\rm mag. mom.}=0.7$,
$\chi^2_{\rm SLD}=0.3$, with total $\chi^2/{\rm DOF}=1/5$.
If one considers $A_0$ and $A_1$ as parameters then $\chi^2/{\rm DOF}=1/3$.
The values of the sea parameters, $\mu_u$, and $\mu_s$ are given at the end of
Table~\ref{tabla7n}.

Comparison of this 7 parameter fit with the earlier 6 parameter fits for
magnetic moments alone reveals:

1) The new values of $\mu_q$'s are close to the earlier ones.
But, the fit has the nice feature that $m_u=m_d$ and $m_u/m_s$ is closer to 0.6.

2) The scalar sea is described by the 2 parameters $a({\bf 8_F})$ and
$a({\bf 10})$ in all the three cases with values which are quite close.
One finds the range of values
$a({\bf 8_F})\approx-0.1536$ to $-0.1465$ and $a({\bf 10})\approx 0.4941$ to $0.5130$.

3) The vector sea in the 7 parameter fit is described by 3 parameters
$b_0=0.3060$, $b({\bf 8_F})=-0.3296$ and $b({\bf \bar{10}})=0.2442$.
In contrast, the six parameter fits had only one parameter $b_0=0.4779$ or
$b({\bf 8_F})=0.5272$.

4) $SU(3)$ breaking effects in the wavefunction in the fit of
Table~\ref{tabla7n} are due to the vector sea parameter $b({\bf \bar{10}})$ in
addition to the scalar sea parameter $a({\bf 10})$ of the earlier fits.

5) The values of $(\Delta q)^p$ are similar except that they fit the
experimental value of $G_A/G_V(n\to p)=(\Delta u)^p-(\Delta d)^p$ precisely now.
Also, the direct sea contribution $G^{(s)}_A(\Delta S=1)$ with $A_1=-1$ is
necessary to fit the $\Sigma^-\to n$ decay.

In summary, at the expense of an extra parameter overall one obtains a better
fit to magnetic moment data than before as well as fit the known SLD data {\em
using experimental errors throughout}.

\subsection{Fits with theoretical errors of ${\bf 0.1\mu_N}$ for magnetic
moments}
\label{vb}

We consider such fits because all the fits in the literature (unlike our fits
above) add an arbitrary theoretical error.
The motivation for adding this error is that all magnetic moments are treated
\lq\lq democratically". 
Otherwise, the extremely accurately measured $\mu_p$ and $\mu_n$ act as inputs
to a minimum $\chi^2$-fit.
We add a theoretical error of $0.1\mu_N$ in quadratures to the experimental
errors for all the magnetic moment data.
This is a popular choice\cite{karl,casu}.
An error of $0.1\mu_N$ is fairly large (compared to the actual experimental
errors) and facilitates a good fit with a few parameters only.
This is true in our model also!
A 3 parameter fit with inputs $m_u=m_d=0.6m_s$ and $A_0=A_1=0$ is given in
column 4 of Table~\ref{tabla7n}.
In this fit, the scalar and vector seas are described by one parameter each,
namely $a({\bf 10})$ and $b({\bf 8_D})$ respectively.

How does our fit compare to other fits with $0.1\mu_N$ theoretical error?
We give a comparison with the most recent fits\cite{casu} refered to as CS
below.
Unlike us the model of CS does not fit $\mu_{\Sigma^0\Lambda}$ as it is not
clear how to include it in their picture of 3-quark correlation within a baryon.
For magnetic moments alone our fit give
$\chi^2_{\rm mag. mom.}/{\rm DOF}=3.8/5$ compared to $4.4/4$ for Models AII and
AIII of CS, their best fits.
An important difference in their and our model is reflected in the
phenomenological values of $(\Delta q)^B$.
In particular, CS obtain (their Model AIII) $(\Delta u)^p=0.783$,
$(\Delta d)^p=-0.477$, and $(\Delta s)^p=-0.147$.
This is to be contrasted with the fact that our fits yield $(\Delta u)^p=0.964$,
$(\Delta d)^p=-0.296$, and $(\Delta s)^p=0.008$.
Physically, our fits require a very tiny strange-quark content in the nucleon
compared to their and other similar fits\cite{karl,casu}.
Another physical difference is that in our case the valence quarks carry $67\%$
of the proton spin compared to about $16\%$ in Model AIII of CS.

\subsection{Fits with smaller theoretical errors for magnetic moments}
\label{vc}

Since all the magnetic moments, except for $p$ and $n$, are known to three
digits one can treat all of them democratically with a theoretical error of
$0.001\mu_N$.
Using this error in quadratures and the predicted numbers in Column~3 of
Table~\ref{tabla7n}, one finds $\chi^2\simeq 1$ instead of 1.02 for
experimental errors.
An independent fit with theoretical errors of $0.001\mu_N$ gives a $\chi^2=1.0035$.
In fact, fits with theoretical errors of 0.001$\mu_N$ or smaller are
essentially equivalent to the fits using experimental errors (the changes being
only in the fourth decimal place or after).
The situation changes however when larger theoretical errors like $0.01\mu_N$
are used. 
We now turn to implications for the spin distribution of our model. 

\section{Spin distributions}
\label{vi}

The spin distribution, $I_{1B}$, for baryon $B$ is defined as

\begin{equation}
I_{1B}\equiv\int^1_0g_{1B}(x)dx,
\label{catorcen}
\end{equation}

\noindent
where the spin structure function $g_{1B}(x)$ occurs in polarized
electron-baryon scattering.
Experiment\cite{1,smc} gives $I_{1p}=0.126\pm 0.018$ and $I_{1n}=-0.08\pm 0.06$
which are very different from the SQM predictions $I_{1p}=5/18=0.2778$ and
$I_{1n}=0$.
One must note that the EMC experiment gives $I_{1p}$ for
$\langle Q^2\rangle=10.7({\rm GeV}/c)^2$ and this could be very different for
the very low $Q^2\ (\approx 0)$ result predicted by SQM or other theoretical
models.
This could mean that a model which gives value for $I_{1B}$ differeng by 2-3
standard deviations from experiment may be quite acceptable.

In SQM, $I_{1B}$ is given by the expectation value
$I^{(q)}_{1B}\equiv\langle B|\hat{I}^{(q)}_{1}|B\rangle$ where the quark
operator $\hat{I}^{(q)}_{1}=(1/2)\sum_qe^2_q\sigma^q_Z$.
This gives

\begin{equation}
I^{(q)}_{1B}=\frac{1}{18}[4(\Delta u)^B+(\Delta d)^B+(\Delta s)^B].
\label{quincen}
\end{equation}

In our model in addition to the quarks there can be a direct sea contribution
$I^{(s)}_{1B}\equiv\langle B|\hat{I}^{(s)}_{1}|B\rangle$ where by analogy we
take $\hat{I}^{(s)}_{1}=e^2_sS^{(s)}_Z$.
Thus only the charged states in the vector sea will contribute to
$I^{(s)}_{1B}$.
For the nucleons, one obtains

\begin{equation}
I^{(s)}_{1p}=\frac{2}{3N^2_1}
\left(\bar{\beta}'^2_2+\frac{2}{3}\bar{\beta}'^2_3+
\frac{1}{3}\bar{\beta}'^2_4\right),
\qquad
I^{(s)}_{1n}=\frac{2}{3N^2_1}\left(\frac{2}{3}\bar{\beta}'^2_3+
\frac{2}{3}\bar{\beta}'^2_4\right).
\label{dieciseisn}
\end{equation}

\noindent
Putting the two contributions together we have

\begin{equation}
I_{1B}=I^{(q)}_{1B}+B_1I^{(s)}_{1B},
\label{diecisieten}
\end{equation}

\noindent
where $B_1$ determines the strength of the direct sea contribution to the
valence quark contribution.
Since the value of $B_1$ is not known \'a priori and phenomenologically it may
be treated as a parameter.

Using the fit to magnetic moment and SLD data with experimental errors given in
Table~\ref{tabla7n} we can predict $I_{1B}$.
One obtains $I^{(q)}_{1p}=0.205$, $I^{(q)}_{1n}=-0.005$, while,
$I^{(s)}_{1p}=0.044$ and $I^{(s)}_{1n}=0.057$, where we have used
$(\Delta u)^p=0.989$, $(\Delta d)^p=-0.271$, and $(\Delta s)^p=0.009$.
If one keeps only the quark part, that is $B_1=0$, then our $I_{1p}$ is much
lower than the SQM value but still $4\sigma$ higher than experiment.
This may be due to large $\langle Q^2\rangle$ in the experiment.
Another possibility is to invoke the direct sea contribution.
For example, with $B_1=-1$ one obtains $I_{1p}=0.161$ and $I_{1n}=-0.062$ in
good agreement with experiment.

\section{Summary}
\label{vii}

In summary, we have shown that our model of the sea component in spin-1/2
baryons can fit their magnetic moments, weak decay constants $G_A/G_V$ for both
$\Delta S=0$ and 1 semileptonic decays as well as nuclear spin distributions
{\em using experimental errors.}
To accomplish this one has to invoke a direct sea contribution for  $\Delta S=1$
decays and nucleon spin distribution.
The sea was found to be both scalar (spin 0) and vector (spin 1).
Two physical features of our fits are that about $70\%$ of the proton spin
resides with the valence quarks and they give a tiny strange-quark content to the
nucleon.

\acknowledgments

This work was partially supported by CONACyT (M\'exico).

\mediumtext

\begin{table}
\caption{
Contribution to the physical baryon state $B(Y,I,I_3)$ formed out of
$\tilde{B}(Y,I,I_3)$ and flavor octet states $S(Y,I,I_3)$ (see third and
fourth terms in Eq.~(1)).
The core baryon states $\tilde{B}$ denoted by $\tilde{p}$, $\tilde{n}$, etc.
are the normal 3 valence quark states of SQM.
The sea octet states are denoted by $S_{\pi^+}=S(0,1,1)$, etc. as in Eq.~(3).
Further, $(\tilde{N}S_{\pi})_{I,I_3}$, $(\tilde{\Sigma}S_{\bar{K}})_{I,I_3}$,
$(\tilde{\Sigma}S_{\pi})_{I,I_3}$, $\dots$ stand for total $I$, $I_3$
{\em normalized} combinations of $\tilde{N}$ and $S_{\pi}$, etc.
See Table~II for the coefficients $\bar{\beta}_i$, $\beta_i$, $\gamma_i$,
and $\delta_i$.
}~
\label{tabla1}
\begin{tabular}
{
cc
}
$B(Y,I,I_3)$ &
$\tilde{B}(Y,I,I_3)$ and $S(Y,I,I_3)$
\\
\hline
\\
$p$ &
$
\bar{\beta}_1 \tilde{p} S_{\eta} 
+ \bar{\beta}_2 \tilde{\Lambda} S_{K^+} 
+ \bar{\beta}_3 (\tilde{N}S_{\pi})_{1/2,1/2}
+ \bar{\beta}_4 (\tilde{\Sigma}S_{K})_{1/2,1/2}
$
\\
\\
$n$ &
$
\bar{\beta}_1 \tilde{n} S_{\eta} 
+ \bar{\beta}_2 \tilde{\Lambda} S_{K^0} 
+ \bar{\beta}_3 (\tilde{N}S_{\pi})_{1/2,-1/2}
+ \bar{\beta}_4 (\tilde{\Sigma}S_{K})_{1/2,-1/2}
$
\\
\\
$\Xi^0$ &
$
\beta_1 \tilde{\Xi}^0 S_{\eta} 
+ \beta_2 \tilde{\Lambda} S_{\bar{K}^0} 
+ \beta_3 (\tilde{\Xi}S_{\pi})_{1/2,1/2}
+ \beta_4 (\tilde{\Sigma}S_{\bar{K}})_{1/2,1/2}
$
\\
\\
$\Xi^-$ &
$
\beta_1 \tilde{\Xi}^- S_{\eta} 
+ \beta_2 \tilde{\Lambda} S_{\bar{K}^-} 
+ \beta_3 (\tilde{\Xi}S_{\pi})_{1/2,-1/2}
+ \beta_4 (\tilde{\Sigma}S_{\bar{K}})_{1/2,-1/2}
$
\\
\\
$\Sigma^+$ &
$
\gamma_1 \tilde{p} S_{\bar{K}^0} 
+ \gamma_2 \tilde{\Xi}^0 S_{K^+} 
+ \gamma_3 \tilde{\Lambda} S_{\pi^+}
+ \gamma_4 \tilde{\Sigma}^+ S_{\eta}
+ \gamma_5 (\tilde{\Sigma}S_{\pi})_{1,1}
$
\\
\\
$\Sigma^-$ &
$
\gamma_1 \tilde{n} S_{K^-} 
+ \gamma_2 \tilde{\Xi}^- S_{K^0} 
+ \gamma_3 \tilde{\Lambda} S_{\pi^-}
+ \gamma_4 \tilde{\Sigma}^- S_{\eta}
+ \gamma_5 (\tilde{\Sigma}S_{\pi})_{1,-1}
$
\\
\\
$\Sigma^0$ &
$
\gamma_1 (\tilde{N} S_{\bar{K}})_{1,0} 
+ \gamma_2 (\tilde{\Xi} S_{K})_{1,0} 
+ \gamma_3 \tilde{\Lambda} S_{\pi^0}
+ \gamma_4 \tilde{\Sigma}^0 S_{\eta}
+ \gamma_5 (\tilde{\Sigma}S_{\pi})_{1,0}
$
\\
\\
$\Lambda$ &
$
\delta_1 (\tilde{N} S_{\bar{K}})_{0,0} 
+ \delta_2 (\tilde{\Xi} S_{K})_{0,0} 
+ \delta_3 \tilde{\Lambda} S_{\eta}
+ \delta_4 (\tilde{\Sigma}S_{\pi})_{0,0}
$
\\
\end{tabular}
\end{table}

\widetext

\begin{table}
\squeezetable
\caption{
The coefficients $\bar{\beta}_i$, $\beta_i$, $\gamma_i$, and $\delta_i$ in
Table~I expressed in terms of the coefficients $a(N)$,
$N={\bf 1, 8_{F}, 8_{D}, 10, \bar{10}, 27}$, in the $3^{\rm rd}$ term
(from scalar sea) in Eq.~(1).
The corresponding coefficients $\bar{\beta}'_i$, $\beta'_i$, $\gamma'_i$,
and $\delta'_i$ determining the flavor of structure of $4^{\rm th}$ term in
Eq.~(1) can be obtained from $\bar{\beta}_i$, etc. by the replacement
$a(N)\rightarrow b(N)$ (see text).
}~
\label{tabla2}
\begin{tabular}
{
cc
}
\\
$
\bar{\beta}_1=\frac{1}{\sqrt {20}}(3a({\bf 27})-a({\bf 8_D}))+
\frac{1}{2}(a({\bf 8_F})+a({\bf \bar{10}}))
$
&
$
\beta_1=\frac{1}{\sqrt {20}}(3a({\bf 27})-a({\bf 8_D}))-
\frac{1}{2}(a({\bf 8_F})-a({\bf 10}))
$
\\
\\
$
\bar{\beta}_2=\frac{1}{\sqrt {20}}(3a({\bf 27})-a({\bf 8_D}))-
\frac{1}{2}(a({\bf 8_F})+a({\bf \bar{10}}))
$
&
$
\beta_2=\frac{1}{\sqrt {20}}(3a({\bf 27})-a({\bf 8_D}))+
\frac{1}{2}(a({\bf 8_F})-a({\bf 10}))
$
\\
\\
$
\bar{\beta}_3=\frac{1}{\sqrt {20}}(a({\bf 27})+3a({\bf 8_D}))+
\frac{1}{2}(a({\bf 8_F})-a({\bf \bar{10}}))
$
&
$
\beta_3=-\frac{1}{\sqrt {20}}(a({\bf 27})+3a({\bf 8_D}))+
\frac{1}{2}(a({\bf 8_F})+a({\bf 10}))
$
\\
\\
$
\bar{\beta}_4 =-\frac{1}{\sqrt {20}}(a({\bf 27})+3a({\bf 8_D}))+
\frac{1}{2}(a({\bf 8_F})-a({\bf \bar{10}}))
$
&
$
\beta_4=\frac{1}{\sqrt{20}}(a({\bf 27})+3a({\bf 8_D}))+
\frac{1}{2}(a({\bf 8_F})+a({\bf 10}))
$
\\
\\
$
\gamma_1=\frac{1}{\sqrt{10}}(\sqrt{2}a({\bf 27})-\sqrt{3}a({\bf 8_D}))+
\frac{1}{\sqrt{6}}(a({\bf 8_F})-a({\bf 10})+a({\bf\bar{10}}))
$
&
$
\delta_1=\frac{1}{\sqrt{20}}(\sqrt{3}a({\bf 27})+\sqrt{2}a({\bf 8_D}))+
\frac{1}{2}(\sqrt{2}a({\bf 8_F})+a({\bf 1}))
$
\\
\\
$
\gamma_2=\frac{1}{\sqrt{10}}(\sqrt{2}a({\bf 27})-\sqrt{3}a({\bf 8_D}))-
\frac{1}{\sqrt{6}}(a({\bf 8_F})-a({\bf 10})+a({\bf\bar{10}}))
$
&
$
\delta_2=-\frac{1}{\sqrt{20}}(\sqrt{3}a({\bf 27})+\sqrt{2}a({\bf 8_D}))+
\frac{1}{2}(\sqrt{2}a({\bf 8_F})-a({\bf 1}))
$
\\
\\
$
\gamma_3=\frac{1}{\sqrt{10}}(\sqrt{3}a({\bf 27})+\sqrt{2}a({\bf 8_D}))-
\frac{1}{2}(a({\bf 10})+a({\bf\bar{10}}))
$
&
$
\delta_3=\frac{3\sqrt{3}}{\sqrt{40}}a({\bf 27})-
\frac{1}{\sqrt{5}}a({\bf 8_D})-
\frac{\sqrt{2}}{4}a({\bf 1})
$
\\
\\
$
\gamma_4=\frac{1}{\sqrt{10}}(\sqrt{3}a({\bf 27})+\sqrt{2}a({\bf 8_D}))+
\frac{1}{2}(a({\bf 10})+a({\bf\bar{10}}))
$
&
$
\delta_4=-\frac{1}{\sqrt{40}}a({\bf 27})-\sqrt{\frac{3}{5}}a({\bf 8_D})+
\frac{\sqrt{6}}{4}a({\bf 1})
$
\\
\\
$
\gamma_5=\frac{1}{\sqrt{6}}(2a({\bf 8_F})+a({\bf 10})-a({\bf\bar{10}}))
$
&
\\
\end{tabular}
\end{table}

\begin{table}
\squeezetable
\caption{
$(\Delta q)^B$ defined in Eq.~(8) for physical baryon $B$ given by general
wavefunction in Eq.~(1).
The normalizations $N_1$, $N_2$, $N_3$, and $N_4$ are given in Eqs.~(5).
The $(\Delta q)^{\Sigma^0\Lambda}$ for the $\Sigma^0\rightarrow\Lambda$
transition magnetic moment is also given.
}~
\label{tabla3}
\begin{tabular}
{
c
}
\\
$
(\Delta u)^p=
\frac{1}{3N^2_1}
[
4(1-\frac{1}{3}b^2_0)
+
(
4\bar{\beta}^2_1+\frac{2}{3}\bar{\beta}^2_3+\frac{10}{3}\bar{\beta}^2_4
-2\bar{\beta}_2\bar{\beta}_4
)
-\frac{1}{3}
(
\bar{\beta}_i\rightarrow\bar{\beta}'_i
)
]
$
\\
\\
$
(\Delta d)^p=
\frac{1}{3N^2_1}
[
-(1-\frac{1}{3}b^2_0)
+
(
-\bar{\beta}^2_1+\frac{7}{3}\bar{\beta}^2_3+\frac{2}{3}\bar{\beta}^2_4
+2\bar{\beta}_2\bar{\beta}_4
)
-\frac{1}{3}
(
\bar{\beta}_i\rightarrow\bar{\beta}'_i
)
]
\ \ \ \ \ \ 
(\Delta s)^p=
\frac{1}{3N^2_1}
[
(
3\bar{\beta}^2_2-\bar{\beta}^2_4
)
-\frac{1}{3}
(
\bar{\beta}_i\rightarrow\bar{\beta}'_i
)
]
$
\\
\\
$
(\Delta u)^n=(\Delta d)^p
\ \ \ \ \ \ 
(\Delta d)^n=(\Delta u)^p
\ \ \ \ \ \ 
(\Delta s)^n=(\Delta s)^p
$
\\
\\
$
(\Delta u)^{\Xi^0}=
\frac{1}{3N^2_2}
[
-(1-\frac{1}{3}b^2_0)
+
(
-\beta^2_1-\frac{1}{3}\beta^2_3+\frac{10}{3}\beta^2_4
-2\beta_2\beta_4
)
-\frac{1}{3}
(
\beta_i\rightarrow\beta'_i
)
]
$
\\
\\
$
(\Delta d)^{\Xi^0}=
\frac{1}{3N^2_2}
[
(
-\frac{2}{3}\beta^2_3+\frac{2}{3}\beta^2_4
+2\beta_2\beta_4
)
-\frac{1}{3}
(
\beta_i\rightarrow\beta'_i
)
]
\ \ \ \ \ \ 
(\Delta s)^{\Xi^0}=
\frac{1}{3N^2_2}
[
4(1-\frac{1}{3}b^2_0)
+
(
4\beta^2_1+3\beta^2_2+4\beta^2_3-\beta^2_4
)
-\frac{1}{3}
(
\beta_i\rightarrow\beta'_i
)
]
$
\\
\\
$
(\Delta u)^{\Xi^-}=(\Delta d)^{\Xi^0}
\ \ \ \ \ \ 
(\Delta d)^{\Xi^-}=(\Delta u)^{\Xi^0}
\ \ \ \ \ \ 
(\Delta s)^{\Xi^-}=(\Delta s)^{\Xi^0}
$
\\
\\
$
(\Delta u)^{\Sigma^+}=
\frac{1}{3N^2_3}
[
4(1-\frac{1}{3}b^2_0)
+
(
4\gamma^2_1-\gamma^2_2+4\gamma^2_4+3\gamma^2_5
-\sqrt{6}\gamma_3\gamma_5
)
-\frac{1}{3}
(
\gamma_i\rightarrow\gamma'_i
)
]
$
\\
\\
$
(\Delta d)^{\Sigma^+}=
\frac{1}{3N^2_3}
[
(
-\gamma^2_1+\gamma^2_5
+\sqrt{6}\gamma_3\gamma_5
)
-\frac{1}{3}
(
\gamma_i\rightarrow\gamma'_i
)
]
\ \ \ \ \ \ 
(\Delta s)^{\Sigma^+}=
\frac{1}{3N^2_3}
[
-(1-\frac{1}{3}b^2_0)
+
(
4\gamma^2_2+3\gamma^2_3-\gamma^2_4-\gamma^2_5
)
-\frac{1}{3}
(
\gamma_i\rightarrow\gamma'_i
)
]
$
\\
\\
$
(\Delta u)^{\Sigma^-}=(\Delta d)^{\Sigma^+}
\ \ \ \ \ \ 
(\Delta d)^{\Sigma^-}=(\Delta u)^{\Sigma^+}
\ \ \ \ \ \ 
(\Delta s)^{\Sigma^-}=(\Delta s)^{\Sigma^+}
$
\\
\\
$
(\Delta u)^{\Sigma^0}=
\frac{1}{2}[(\Delta u)^{\Sigma^+}+(\Delta u)^{\Sigma^-}]
\ \ \ \ \ \ 
(\Delta d)^{\Sigma^0}=(\Delta u)^{\Sigma^0}
\ \ \ \ \ \ 
(\Delta s)^{\Sigma^0}=(\Delta s)^{\Sigma^+}
$
\\
\\
$
(\Delta u)^{\Lambda}=
\frac{1}{3N^2_4}
[
(
\frac{3}{2}\delta^2_1-\frac{1}{2}\delta^2_2+2\delta^2_4
)
-\frac{1}{3}
(
\delta_i\rightarrow\delta'_i
)
]
$
\\
\\
$
(\Delta d)^{\Lambda}=(\Delta u)^{\Lambda}
\ \ \ \ \ \ \ \ \ 
(\Delta s)^{\Lambda}=
\frac{1}{3N^2_4}
[
3(1-\frac{1}{3}b^2_0)
+
(
4\delta^2_2+3\delta^2_3-\delta^2_4
)
-\frac{1}{3}
(
\delta_i\rightarrow\delta'_i
)
]
$
\\
\\
$
(\Delta u)^{\Sigma^0\Lambda}=
\frac{1}{N_3N_4}
[
\frac{1}{\sqrt{3}}(1-\frac{1}{3}b^2_0)
+
(
\frac{1}{\sqrt{3}}\gamma_4\delta_3-\frac{1}{3}\gamma_3\delta_4
+
\frac{5}{6}\gamma_1\delta_1-\frac{1}{6}\gamma_2\delta_2
+\frac{4}{3\sqrt{6}}\gamma_5\delta_4
)
-\frac{1}{3}
(
\gamma_i,\delta_i\rightarrow\gamma'_i,\delta'_i
)
]
$
\\
\\
$
(\Delta d)^{\Sigma^0\Lambda}=-(\Delta u)^{\Sigma^0\Lambda}
\ \ \ \ \ \ \ \ \ \ \ \ 
(\Delta s)^{\Sigma^0\Lambda}=0
$
\\
\end{tabular}
\end{table}

\begin{table}
\caption{
$\Delta S=0$ semileptonic decays.
Total $G_{V}(B\to B')$ with $G^{(q)}_{A}(B\to B')$ and
$G^{(s)}_{A}(B\to B')$, the quark and direct sea contributions to total
$G_{A}(B\to B')$.
}~
\label{tabla4n}
\begin{tabular}
{
cccc
}
Decay &
$G_{V}(B\to B')$ &
$G^{(q)}_{A}(B\to B')$ &
$G^{(s)}_{A}(B\to B')$
\\
\hline
\\
$n\to p$ &
1 &
$(\Delta d)^n-(\Delta d)^p$ &
$\frac{4}{3N^2_1}(\bar{\beta}'^2_2+\frac{4}{3}\bar{\beta}'^2_3-
\frac{1}{3}\bar{\beta}'^2_4)$
\\
\\
$\Xi^-\to \Xi^0$ &
1 &
$(\Delta d)^{\Xi^-}-(\Delta d)^{\Xi^0}$ &
$\frac{4}{3N^2_2}(\beta'^2_2+\frac{4}{3}\beta'^2_3-
\frac{1}{3}\beta'^2_4)$
\\
\\
$\Sigma^-\to \Sigma^0$ &
$\sqrt{2}$ &
$\sqrt{2}[(\Delta d)^{\Sigma^-}-(\Delta d)^{\Sigma^0}]$ &
$\frac{2\sqrt{2}}{3N^2_3}(\gamma'^2_1+\gamma'^2_2+2\gamma'^2_3+
\gamma'^2_5)$
\\
\\
$\Sigma^0\to \Sigma^+$ &
$\sqrt{2}$ &
$G^{(q)}_{A}(\Sigma^-\to \Sigma^0)$ &
$G^{(s)}_{A}(\Sigma^-\to \Sigma^0)$
\\
\\
$\Sigma^+\to \Lambda$ &
0 &
$\sqrt{2}(\Delta d)^{\Sigma^0\Lambda}$ &
$\frac{4}{3N_3N_4}[\frac{1}{\sqrt{2}}(\gamma'_1\delta'_1+
\gamma'_2\delta'_2)+\frac{2}{\sqrt{3}}\gamma'_5\delta'_4]$
\\
\\
$\Sigma^-\to \Lambda$ &
0 &
$-G^{(q)}_{A}(\Sigma^+\to \Lambda)$ &
$-G^{(s)}_{A}(\Sigma^+\to \Lambda)$
\\
\end{tabular}
\end{table}

\begin{table}
\caption{
Total $G_{V}(B\to B')$ for $\Delta S=1$ semileptonic decays.
}~
\label{tabla5n}
\begin{tabular}
{
cc
}
Decay &
$G_{V}(B\to B')$
\\
\hline
\\
$\Lambda\to p$ &
$
\sqrt{\frac{3}{2}}\frac{1}{N_1N_4}
[
1+b^2_0+2\sqrt{\frac{2}{3}}a^2({\bf 27})+
a^2({\bf 8_D})+a^2({\bf 8_F})+(a(N)\to b(N))
]
$
\\
\\
$\Sigma^-\to n$ &
$
\frac{1}{N_1N_3}
[
1+b^2_0+2a^2({\bf\bar{10}})+2\sqrt{\frac{2}{3}}a^2({\bf 27})+
a^2({\bf 8_D})+a^2({\bf 8_F})+(a(N)\to b(N))
]
$
\\
\\
$\Sigma^0\to p$ &
$
\frac{1}{\sqrt{2}}G_{V}(\Sigma^-\to n)
$
\\
\\
$\Xi^-\to\Lambda$ &
$
\sqrt{\frac{3}{2}}\frac{1}{N_2N_4}
[
1+b^2_0+2\sqrt{\frac{2}{3}}a^2({\bf 27})+
a^2({\bf 8_D})+a^2({\bf 8_F})+(a(N)\to b(N))
]
$
\\
\\
$\Xi^-\to\Sigma^0$ &
$
\frac{1}{\sqrt{2}}\frac{1}{N_3N_4}
[
1+b^2_0+2a^2({\bf 10})+2\sqrt{\frac{2}{3}}a^2({\bf 27})+
a^2({\bf 8_D})+a^2({\bf 8_F})+(a(N)\to b(N))
]
$
\\
\\
$\Xi^0\to\Sigma^+$ &
$
\sqrt{2}G_{V}(\Xi^-\to\Sigma^0)
$
\\
\end{tabular}
\end{table}

\begin{table}
\squeezetable
\caption{
$G^{(q)}_{A}(B\to B')$ and $G^{(s)}_{A}(B\to B')$ the quark and direct sea
contributions to total $G_{A}(B\to B')$ for $\Delta S=1$ semileptonic decays.
}~
\label{tabla6n}
\begin{tabular}
{
cl
}
Decay &
\multicolumn{1}{c}{$G^{(q)}_{A}(B\to B')$ and $G^{(s)}_{A}(B\to B')$}
\\
\hline
\\
$\Lambda\to p$ &
$
G^{(q)}_{A}=
\sqrt{\frac{3}{2}}\frac{1}{N_1N_4}
[
1-\frac{1}{3}b^2_0+\delta_3\bar{\beta}_1+
\frac{1}{3\sqrt{2}}\delta_2(-\bar{\beta}_2+5\bar{\beta}_4)+
\frac{1}{3\sqrt{3}}\delta_4\bar{\beta}_3
-\frac{1}{3}(\delta_i,\bar{\beta}_i\to\delta_i',\bar{\beta}_i')
]
$
\\
&
$
G^{(s)}_{A}=
2\sqrt{\frac{2}{3}}\frac{1}{N_1N_4}
[
\frac{1}{\sqrt{2}}\delta_1'(\bar{\beta}_1'+\bar{\beta}_3')+
\delta_3'\bar{\beta}_2'+\frac{1}{\sqrt{3}}\delta_4'\bar{\beta}_4'
]
$
\\
\\
$\Sigma^-\to n$ &
$
G^{(q)}_{A}=
-\frac{1}{3N_1N_3}
[
1-\frac{1}{3}b^2_0-\frac{1}{\sqrt{6}}\gamma_2(3\bar{\beta}_2+5\bar{\beta}_4)-
3\gamma_3\bar{\beta}_3+\gamma_4\bar{\beta}_1+
\sqrt{\frac{2}{3}}\gamma_5\bar{\beta}_3
-\frac{1}{3}(\gamma_i,\bar{\beta}_i\to\gamma_i',\bar{\beta}_i')
]
$
\\
&
$
G^{(s)}_{A}=
\frac{4}{3}\frac{1}{N_1N_3}
[
\frac{1}{\sqrt{6}}\gamma'_1(3\bar{\beta}_1'-\bar{\beta}_3')+
\gamma_3'\bar{\beta}_2'-\gamma_4'\bar{\beta}_4'+
\sqrt{\frac{2}{3}}\gamma'_5\bar{\beta}'_4
]
$
\\
\\
$\Sigma^0\to p$ &
$
G^{(q)}_{A}=\frac{1}{\sqrt{2}}G^{(q)}_{A}(\Sigma^-\to n)
$
\\
&
$
G^{(s)}_{A}=\frac{1}{\sqrt{2}}G^{(s)}_{A}(\Sigma^-\to n)
$
\\
\\
$\Xi^-\to\Lambda$ &
$
G^{(q)}_{A}=
\frac{1}{\sqrt{6}}\frac{1}{N_2N_4}
[
1-\frac{1}{3}b^2_0+
\frac{1}{\sqrt{2}}\delta_1(3\beta_2-\beta_4)+
\delta_3\beta_1+\frac{5}{\sqrt{3}}\delta_4\beta_3
-\frac{1}{3}(\delta_i,\beta_i\to\delta_i',\beta'_i)
]
$
\\
&
$
G^{(s)}_{A}=
2\sqrt{\frac{2}{3}}\frac{1}{N_2N_4}
[
\frac{1}{\sqrt{2}}\delta'_2(\beta'_3-\beta'_1)+
\delta'_3\beta'_2-\frac{1}{\sqrt{3}}\delta'_4\beta'_4
]
$
\\
\\
$\Xi^-\to\Sigma^0$ &
$
G^{(q)}_{A}=
\frac{5}{3\sqrt{2}}\frac{1}{N_3N_4}
[
1-\frac{1}{3}b^2_0+\frac{3}{5}\sqrt{\frac{3}{2}}\gamma_1\beta_2+
\frac{1}{5\sqrt{6}}\gamma_1\beta_4-\frac{1}{5}\gamma_3\beta_3+
\gamma_4\beta_1+\sqrt{\frac{2}{3}}\gamma_5\beta_3
-\frac{1}{3}(\gamma_i,\beta_i\to\gamma'_i,\beta'_i)
]
$
\\
&
$
G^{(s)}_{A}=
\frac{2\sqrt{2}}{3}\frac{1}{N_3N_4}
[
\frac{1}{\sqrt{6}}\gamma'_2(3\beta'_1+\beta'_3)+
\gamma'_3\beta'_2+\gamma'_4\beta'_4+
\sqrt{\frac{2}{3}}\gamma'_5\beta'_4
]
$
\\
\\
$\Xi^0\to\Sigma^+$ &
$
G^{(q)}_{A}=\sqrt{2}G^{(q)}_{A}(\Xi^-\to\Sigma^0)
$
\\
&
$
G^{(s)}_{A}=\sqrt{2}G^{(s)}_{A}(\Xi^-\to\Sigma^0)
$
\\
\end{tabular}
\end{table}

\begin{table}
\squeezetable
\caption{
Combined fits to the semileptonic decay and magnetic moment data.
All the magnetic moment values are given in nuclear magnetons, $\mu_N$.
a) Fit with experimental errors for all data, see Sec.~\ref{va} (column 3).
b) Fit with theoretical errors of $0.1\mu_N$ added in quadratures for magnetic
moments, see Sec.~\ref{vb} (column 4).
}~
\label{tabla7n}
\begin{tabular}
{
l
r@{\,$\pm$\,}l
d
d
}
&
\multicolumn{2}{c}{Data\cite{pdg}} &
a) Experimental errors &
b) Theoretical errors
\\
\hline
$\mu(p)$ &
2.79284739 & $6\times 10^{-8}$ &
2.79284739 &
2.79239
\\
$\mu(n)$ &
$-$1.9130428 & $5\times 10^{-7}$ &
$-$1.9130428 &
$-$1.96330
\\
$\mu(\Lambda)$ &
$-$0.613 & 0.004 &
$-$0.613 &
$-$0.608
\\
$\mu(\Sigma^+)$ &
2.458 & 0.010 &
2.458 &
2.538
\\
$\mu(\Sigma^0)$ &
\multicolumn{2}{c}{--------} &
0.6396 &
0.7186
\\
$\mu(\Sigma^-)$ &
$-$1.160 & 0.025 &
$-$1.179 &
$-$1.101
\\
$\mu(\Xi^0)$ &
$-$1.250 & 0.014 &
$-$1.251 &
$-$1.151
\\
$\mu(\Xi^-)$ &
$-$0.6507 & 0.0025 &
$-$0.6506 &
$-$0.5331
\\
$|\mu(\Sigma\Lambda)|$ &
1.61 & 0.08 &
1.59 &
1.55
\\
\\
$G_A/G_V(n\to p)$ &
1.2601 & 0.0025 &
1.2599 &
1.2598
\\
$G_A/G_V(\Lambda\to p)$ &
0.718 & 0.015 &
0.719 &
0.739
\\
$G_A/G_V(\Sigma^-\to n)$ &
$-$0.340 & 0.017 &
$-$0.338 &
$-$0.304
\\
$G_A/G_V(\Xi^-\to \Lambda)$ &
0.25 & 0.05 &
0.22 &
0.22
\\
\\
$\chi^2$/DOF &
\multicolumn{2}{c}{--------} &
\multicolumn{1}{l}{\qquad\qquad 1.02/5} &
\multicolumn{1}{l}{\qquad\qquad 10.70/9}
\\
\\
Inputs &
\multicolumn{2}{c}{--------} &
\multicolumn{1}{l}{\qquad\qquad $m_d=m_u$} &
\multicolumn{1}{l}{\qquad\qquad $m_d=m_u$}
\\
&
\multicolumn{2}{c}{} &
\multicolumn{1}{l}{\qquad\qquad $A_0=0$} &
\multicolumn{1}{l}{\qquad\qquad $m_s=(5/3)m_u$}
\\
&
\multicolumn{2}{c}{} &
\multicolumn{1}{l}{\qquad\qquad $A_1=-1$} &
\multicolumn{1}{l}{\qquad\qquad $A_0=A_1=0$}
\\
\\
Fitted &
\multicolumn{2}{c}{--------} &
\multicolumn{1}{l}{\qquad\qquad $\mu_u=2.4900$} &
\multicolumn{1}{l}{\qquad\qquad $\mu_u=2.5166$}
\\
parameters &
\multicolumn{2}{c}{} &
\multicolumn{1}{l}{\qquad\qquad $\mu_s=-0.7785$} &
\multicolumn{1}{l}{\qquad\qquad $a({\bf 10})=0.5280$}
\\
&
\multicolumn{2}{c}{} &
\multicolumn{1}{l}{\qquad\qquad $a({\bf 8_F})=-0.1465$} &
\multicolumn{1}{l}{\qquad\qquad $b({\bf 8_D})=0.5658$}
\\
&
\multicolumn{2}{c}{} &
\multicolumn{1}{l}{\qquad\qquad $a({\bf 10})=0.5130$} &
\\
&
\multicolumn{2}{c}{} &
\multicolumn{1}{l}{\qquad\qquad $b_0=0.3060$} &
\\
&
\multicolumn{2}{c}{} &
\multicolumn{1}{l}{\qquad\qquad $b({\bf 8_F})=-0.3296$} &
\\
&
\multicolumn{2}{c}{} &
\multicolumn{1}{l}{\qquad\qquad $b({\bf\bar{10}})=0.2442$} &
\\
\end{tabular}
\end{table}

\end{document}